\documentclass[showpacs,prb,preprint]{revtex4}

\usepackage{graphicx}
\usepackage{dcolumn}
\usepackage{amsmath}

\usepackage[T1]{fontenc}
\usepackage{ae}

\begin{document}
%
%
\title[Spin polarization of the $L$-gap surface states on
Au(111)]{Spin polarization of the $L$-gap surface states on Au(111)}

\author{J. Henk}
\email[Corresponding author. Electronic address:\ ]{henk@mpi-halle.de}
\author{A. Ernst}
\author{P. Bruno}
\affiliation{Max-Planck-Institut f\"ur Mikrostrukturphysik, Weinberg~2,
  D-06120 Halle (Saale), Germany}

\date{\today}

\begin{abstract}
  The electron spin polarization (ESP) of the $L$-gap surface states
  on Au(111) is investigated theoretically by means of
  first-principles electronic-structure and photoemission
  calculations. The surface states show a large spin-orbit induced
  in-plane ESP which is perpendicular to the in-plane wavevector, in
  close analogy to a two-dimensional electron gas with Rashba
  spin-orbit interaction.  The surface corrugation leads to a small
  ESP component normal to the surface, being not reported so far.  The
  surface-states ESP can be probed qualitatively and quantitatively by
  spin- and angle-resolved photoelectron spectroscopy, provided that
  the initial-state ESP is retained in the photoemission process and
  not obscured by spin-orbit induced polarization effects.
  Relativistic photoemission calculations provide detailed information
  on what photoemission set-ups allow to conclude from the
  photoelectron ESP on that of the surface states.
\end{abstract}


\maketitle

%
%
\section{Introduction}
Spin-orbit coupling (SOC) is one of the fundamental effects in
condensed-matter physics. It manifests itself in removing degeneracies
in the electronic structure (spin-orbit induced band gaps) which, for
example, leads to the magnetic anisotropy in magnetic systems.
Besides electronic states in the bulk, surface states can become split
by SOC as well. This was shown in a pioneering photoemission
investigation by LaShell and coworkers: they found that the $L$-gap
surface states on Au(111) are split (in binding energy and in-plane
wavevector $\vec{k}_{\parallel}$) and attributed this effect correctly
to SOC\@.~\cite{LaShell96} Later, Petersen and Hedeg{\aa}rd confirmed
this explanation by means of tight-binding
calculations.~\cite{Petersen00} These Shockley surface states are
located in a bulk-band gap which opens up along the $\Gamma$--L
direction (i.\,e., along [111]). Being derived from $sp$-bulk states,
they show almost perfect free-electron dispersion.~\cite{Courths01}
Since the splitting is also present in the other noble metals,
comparative studies on the $L$-gap surface states in Cu, Ag, and Au
were performed by H\"ufner's group using high-resolution
photoemission~\cite{Nicolay00,Reinert01,Nicolay02} (for topical
reviews, see Ref.~\onlinecite{Reinert03} and especially Section~8.2 in
Ref.~\onlinecite{Huefner03}). The photoemission results for Au were
further corroborated experimentally by Fujita \textit{et al.}  using
Fourier-transform scanning tunneling microscopy.~\cite{Fujita99} For
hydrogen-covered W(110) surfaces, the spin-orbit splitting of similar
surface states was also found by angle-resolved
photoemission,~\cite{Rotenberg98} their predicted spin polarization
being confirmed recently.~\cite{Hochstrasser02}

The $L$-gap surface states can be closely related to electronic states
of a two-dimensional electron gas (2DEG) in semiconductor
heterostructures. In the latter, the asymmetry in direction normal to
the semiconductor interface results in the so-called Rashba spin-orbit
interaction.~\cite{Rashba60a,Bychkov84a} In case of the Au surface,
the asymmetry is brought about by the surface potential, in particular
by the surface barrier (i.\,e., a vacuum-solid interface). Therefore,
the $L$-gap surface states can be regarded as being subject to the
Rashba effect, which might render them interesting as a model system
for spintronics.~\cite{Pareek02b,Schliemann03}

In analogy to a 2DEG with Rashba interaction, the spin polarization of
the $L$-gap surface states is assumed to lie within the surface plane
and to be perpendicular to the in-plane wavevector
$\vec{k}_{\parallel}$.~\cite{LaShell96} Further, the split surface
states should show opposite spin polarization. Although the $L$-gap
surface states exhibit spin-orbit induced properties \textit{par
  excellence}, their spin polarization was to our knowledge not
investigated in detail, neither theoretically, nor experimentally.
Probed in a $\vec{k}$-resolved manner, for example by spin- and
angle-resolved photoelectron spectroscopy (SPARPES),~\cite{Huefner03}
their properties should show up unobscured by bulk transitions because
they are located in a bulk-band gap. However, the spin polarization of
the initial state (photohole) is not necessarily that of the
photoelectron, in particular if spin-orbit coupling is strong ($Z =
79$ for gold).  Therefore, the interpretation of spin-resolved
photoemission spectra can become complicated due to the various
spin-polarization effects (SPE's; for atoms, see
Ref.~\onlinecite{Kessler85}). Or stated differently: on one hand, SOC
produces the splitting and the spin polarization of the surface
states. On the other hand, it may prevent to probe the latter by means
of SPARPES due to the SPE's.

The purpose of the present Paper is twofold. First, \textit{ab-initio}
calculations provide detailed information on the properties of the
$L$-gap surface states, in particular on their dispersion and spin
polarization. These results are compared to those for a
two-dimensional electron gas with Rashba spin-orbit interaction.
Second, we address the question whether and how the surface-state spin
polarization can be probed by SPARPES\@. State-of-the-art
photoemission calculations for a variety of set-ups show how the SPE's
affect the photoelectron spin polarization. Former studies of the
photoelectron spin polarization from nonmagnetic surfaces were
performed for normal emission.  For Au(111), the photoelectrons are to
be detected in off-normal emission (due to the dispersion of the
surface states) which leads for some set-ups to distinguished
spin-dependent photoelectron diffraction effects.  Questions to be
answered comprise degree and orientation of the spin polarization as
well its dependence on the wavevector $\vec{k}_{\parallel}$.

This Paper is organized as follows. In Section~\ref{sec:theoretical},
theoretical aspects and relevant details of the computations are
presented. Section~\ref{sec:discussion} focuses first on the analogy
between the electronic states in a 2DEG (\ref{sec:relation-l-gap}) and
the $L$-gap surface states. The properties of the surfaces states are
discussed in \ref{sec:dispersion.esp}, in particular the dispersion
(\ref{sec:dispersion}) and the spin polarization (\ref{sec:esp.ef}).
The theoretical photoemission results are eventually presented in
\ref{sec:probing}.

\section{Computational aspects}
\label{sec:theoretical}
\subsection{The Au(111) surface}
\begin{figure}
  \centering
  \includegraphics[scale = 1.5]{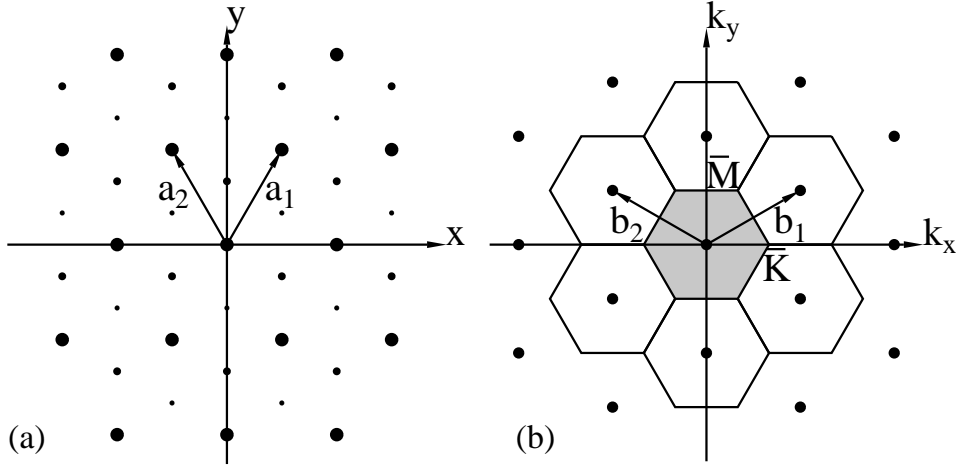}
  \caption{The Au(111) surface. (a) Top view onto the first three
    surface layers (first, second, and third layer: large,
    medium-sized, and small filled circles, respectively).
    $\vec{a}_{1}$ and $\vec{a}_{2}$ are the basis vectors of the
    direct lattice. The $z$ axis points towards the bulk. (b)
    Two-dimensional reciprocal lattice with basis vectors
    $\vec{b}_{1}$ and $\vec{b}_{2}$. The first Brillouin zone is
    marked grey. The two representative symmetry points
    $\overline{\mathrm{K}}$ and $\overline{\mathrm{M}}$ mark a corner
    [$\vec{k}_{\parallel}(\overline{\mathrm{K}}) = (\vec{b}_{1} -
    \vec{b}_{2}) / 2$] and a center
    [$\vec{k}_{\parallel}(\overline{\mathrm{M}}) = (\vec{b}_{1} +
    \vec{b}_{2}) / 2$] of the Brillouin-zone boundary, respectively.}
  \label{fig:geometry}
\end{figure}

The Au(111) surface is shown schematically in
Fig.~\ref{fig:geometry}a.
The $x$ axis corresponds to the crystallographic $[1\overline{1}0]$
direction, whereas the $y$ axis is along
$[\overline{1}\overline{1}2]$.

In the present work, only the $1 \times 1$ unit cell is considered.
The so-called herringbone reconstruction with a $22 \times \sqrt{3}$
unit cell will not be addressed.~\cite{Chen98b} The main effect of
this surface modification is a modulation of the surface-state
photoemission intensities due to backfolding (surface
\textit{umklapp}).

\subsection{Ab-initio calculations}
\label{sec:ab-init-calc}
The electronic structure of the Au(111) surface was computed from
first-principles using the local-density approximation (LDA) of
density-functional theory with Perdew-Wang exchange-correlation
potential.~\cite{Perdew92} The Korringa-Kohn-Rostoker (KKR) method was
applied to semi-infinite systems, hence avoiding slab geometries.

Within our KKR scheme (for details, see Ref.~\onlinecite{Ernst00b}),
we computed first the bulk muffin-tin potentials. Subsequently, the
potentials of the outermost $6$ Au surface layers and of $3$ vacuum
layers were calculated self-consistently for the semi-infinite system.
The Au layers were not relaxed (ideal surface) but for the vacuum
layers an outward relaxation of $4~\%$ (compared to the bulk
interlayer distance) was assumed. The latter improved the dispersion
of the surface states in comparison to the experimental findings
significantly (This procedure clearly reveals shortcomings of the
muffin-tin approximation for the potential, the latter being avoided
in full-potential methods). The work function of $5.23~\mathrm{eV}$
agrees well with the experimental value of
$5.31~\mathrm{eV}$.~\cite{Huefner03}

The \textit{ab-initio} calculations provide details of the electronic
structure by means of the layer- and wavevector-resolved spectral
density,
\begin{equation}
  \label{eq:sd}
  N_{l}(E, \vec{k}_{\parallel})
  =
  -\frac{1}{\pi}
  \mathrm{Im}
  \,\mathrm{Tr}
  \,G_{ll}^{+}(E, \vec{k}_{\parallel}),
\end{equation}
where the trace $\mathrm{Tr}$ is over a muffin-tin sphere of layer $l$
and $G_{ll}^{+}(E, \vec{k}_{\parallel})$ is the $+$ side-limit of the
layer-diagonal Green function at energy $E$ and wavevector
$\vec{k}_{\parallel} = (k_{x}, k_{y})$ (Cartesian coordinates are
defined in Fig.~\ref{fig:geometry}).~\cite{Weinberger90} Further
decomposition of $N_{l}$ with respect to spin and angular momentum
gives access to the relevant surface-state properties.

To investigate the effect of spin-orbit coupling on the $L$-gap
surface states, we scaled the SOC strength by interpolating between
the fully relativistic and the scalar-relativistic
case.~\cite{Tamura.private,Ebert96e} Therefore, only SOC is scaled
whereas the other relativistic effects remain unchanged (note that
this is advantageous compared to scaling the velocity of light).  We
would like to note that this scheme applies only for the muffin-tin
spheres, leaving the gradient of the surface potential almost
unaffected.

\subsection{Photoemission calculations}
\label{sec:phot-calc}
The photoemission calculations were performed using the {\tt omni2k}
computer program for electron spectroscopies~\cite{omni2kb} and rely
on the one-step model as being formulated in the spin-polarized
relativistic layer-KKR method.~\cite{Braun96,Henk01c} Therefore,
spin-orbit coupling is included in a natural way by solving the Dirac
equation. This is in particular important because the SOC-induced
photoelectron spin polarization is fully taken into account.  The
self-consistent potentials from the \textit{ab-initio} calculations
serve as input, putting electronic-structure and photoemission results
on equal footing.

The {\tt omni2k} computer program proved to be successful in a number
of investigations (see Ref.~\onlinecite{Feder96} fur further
publications).  In particular, spin-orbit effects from nonmagnetic
surfaces were described quantitatively [For theoretical predictions of
spin-polarization effects (SPE's) with linearly polarized light, see
Refs.~\onlinecite{Tamura87,Tamura91b,Henk94b}; all these effects were
confirmed experimentally by Heinzmann's
group~\cite{Schmiedeskamp88,Irmer92,Irmer96a}], but also the closely
related magnetic dichroism was addressed
correctly.~\cite{Henk96a,Fanelsa96a,Kuch96b,Rampe98} Hence, we expect
that both the photoemission intensities and the spin polarizations
shown in Section~\ref{sec:probing} agree well with future experiments
on Au(111).

The inverse lifetimes of the photohole (at energies close to the Fermi
level) and of the photoelectron (at about $15~\mathrm{eV}$ kinetic
energy) were chosen as $0.015~\mathrm{eV}$ and $1.25~\mathrm{eV}$,
respectively.  The maximum angular momentum was $l_{\mathrm{max}} = 4$
and the sum over layers comprised the first $30$ layers. Metal optics
were taken into account via Fresnel's equations and Snell's law.

In the following, the incident direction of the light is described by
a polar angle $\vartheta_{\mathrm{ph}}$ and an azimuth
$\varphi_{\mathrm{ph}}$. The in-plane component of the photoelectron
wavevector is given by
\begin{equation}\label{eq:kpar}
  \vec{k}_{\parallel}
  =
  \sqrt{2 E_{\mathrm{kin}}}
  \sin\vartheta_{\mathrm{e}}
  \left(
  \begin{array}{c}
    \cos\varphi_{\mathrm{e}} \\
    \sin\varphi_{\mathrm{e}}
  \end{array}
  \right),
\end{equation}
where $E_{\mathrm{kin}}$ is the kinetic energy. For the
$\overline{\Gamma}$--$\overline{\mathrm{M}}$ direction in the
two-dimensional Brillouin zone (2BZ) one has for example
$\varphi_{\mathrm{e}} = 90^{\circ}$, and for
$\overline{\Gamma}$--$\overline{\mathrm{K}}$ $\varphi_{\mathrm{e}} =
0^{\circ}$ (Fig.~\ref{fig:geometry}b).

\section{Discussion and results}
\label{sec:discussion}
\subsection{Rashba spin-orbit interaction in a two-dimensional electron gas} 
\label{sec:relation-l-gap}
Time-reversal symmetry requires for the dispersion relation
$E(\vec{k}_{\parallel}, \tau) = E(-\vec{k}_{\parallel}, -\tau)$, where
$\tau = \uparrow, \downarrow$ is the electron spin.  Inversion
symmetry (which is present in the bulk of cubic lattices) implies
$E(\vec{k}_{\parallel}, \tau) = E(-\vec{k}_{\parallel}, \tau)$.
Combining these relations yields $E(\vec{k}_{\parallel}, \tau) =
E(\vec{k}_{\parallel}, -\tau)$ (Kramers' degeneracy) which states that
the electronic states in the bulk are not spin-polarized.  However,
the presence of a surface breaks the inversion symmetry and, hence,
spin-orbit induced splitting accompanied by a nonzero spin
polarization is permitted. As Petersen and Hedeg{\aa}rd pointed out,
the splitting depends on both the size of the atomic SOC and of the
gradient of the surface potential.~\cite{Petersen00}

Spin-orbit terms linear in the wavevector $\vec{k}$ occur in the
Hamiltonian due to a symmetry reduction of the system
(heterostructure, film, surface) with respect to the corresponding
bulk system (for a review, see Ref.~\onlinecite{Ganichev03a}).
Particularly important is the structural inversion asymmetry which
occurs typically at semiconductor interfaces [e.\,g., in a
two-dimensional electron gas (2DEG)]~\cite{Luo90} but in fact needs
not to be related to the crystal structure.  In this case, the linear
$\vec{k}_{\parallel}$-terms are the so-called Rashba
terms.~\cite{Rashba60a,Bychkov84a} As will be motivated in the
following, there exists a close analogy between the spin-orbit split
electronic states in a 2DEG and the $L$-gap surface states at (111)
surfaces.

The Hamiltonian of free electrons in two dimensions [$xy$ plane,
$\vec{\rho} = (x, y)$] including the Rashba term can be solved by the
\textit{ansatz}
\begin{equation}
  \Psi(\vec{\rho}, \vec{k}_{\parallel})
  \propto
  \exp(\mathrm{i} \vec{k}_{\parallel} \cdot \vec{\rho})
  \left(
    \pi_{\uparrow} \chi^{\uparrow}
    +
    \pi_{\downarrow} \chi^{\downarrow}
  \right)
\end{equation}
for the wavefunctions. The Pauli spinors $\chi^{\uparrow}$ and
$\chi^{\downarrow}$ are quantized along the $z$ axis. The wavevector
$\vec{k}_{\parallel}$ enters the Schr\"odinger equation~\cite{Units}
\begin{equation}
  \label{eq:2}
  \left[
    \frac{1}{2}
    \vec{k}_{\parallel}^{2}
    \boldsymbol{1}
    +
    \gamma_{\mathrm{so}}
    \left(
    \boldsymbol{\sigma}_{x} k_{y}
    -
    \boldsymbol{\sigma}_{y} k_{x}
  \right)
  \right]
  \Psi
  =
  E
  \Psi,
\end{equation}
where bold symbols represent $2\times 2$ matrices, e.\,g., the Pauli
matrices $\boldsymbol{\sigma}_{i}$, $i = x, y, z$. The parameter
$\gamma_{\mathrm{so}}$, which is assumed positive, controls the
strength of the Rashba spin-orbit interaction.

\begin{figure}
  \centering
  \includegraphics[scale = 0.5]{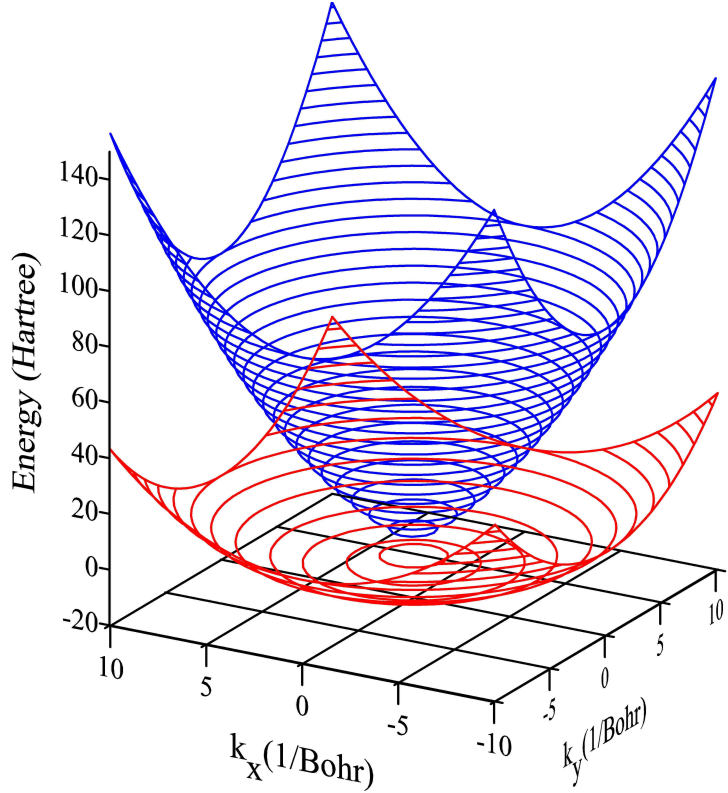}
  \caption{Rashba spin-orbit interaction in a two-dimensional
    electron gas.  The dispersions $E_{\pm}(\vec{k}_{\parallel})$ of
    free electrons are shown for $\gamma_{\mathrm{so}} = 4 /
    \mathrm{Bohr}$, $\vec{k}_{\parallel} = (k_{x}, k_{y})$. The
    `inner' state is represented blue [`$+$' in eq.~(\ref{eq:3})], the
    `outer' red [`$-$' in eq.~(\ref{eq:3})].  Both surfaces touch each
    other at $\vec{k}_{\parallel} = 0$. For a better illustration, the
    Rashba effect is extremely exaggerated (compared to typical
    two-dimensional electron gases).}
  \label{fig:rashba}
\end{figure}

The eigenvalues $E_{\pm}$ of eq.~(\ref{eq:2}) are given by
free-electron parabolae that are shifted in $\vec{k}_{\parallel}$,
\begin{equation}
  \label{eq:3}
  E_{\pm}
  =
  \frac{1}{2}
  k_{\parallel}^{2}
  \pm
  \gamma_{\mathrm{so}}
  |\vec{k}_{\parallel}|.
\end{equation}
The `$+$' solution gives rise to an `inner' paraboloid-like surface
(blue in Fig.~\ref{fig:rashba}), the `$-$' solution to an `outer' one
(red in Fig.~\ref{fig:rashba}).
The associated eigenfunctions $\Psi_{\pm}$ are fully spin-polarized,
as is evident from the spin polarization
\begin{equation}
  \label{eq:pinner}
  \vec{P}_{\pm}(\vec{k}_{\parallel})
  =
  \frac{1}{|\vec{k}_{\parallel}|}
  \left(
    \begin{array}{c}
      \pm k_{y}
      \\
      \mp k_{x}
      \\
      0
    \end{array}
  \right)
  =
  \left(
    \begin{array}{c}
      \pm \sin \varphi_{\mathrm{e}}
      \\
      \mp \cos \varphi_{\mathrm{e}}
      \\
      0
    \end{array}
  \right),
\end{equation}
with $\vec{k}_{\parallel} = |\vec{k}_{\parallel}|
(\cos\varphi_{\mathrm{e}}, \sin\varphi_{\mathrm{e}})$ [cf.\ 
eq.~(\ref{eq:kpar})]. The spin polarization is perpendicular to
$\vec{k}_{\parallel}$, with $P_{+}$ ($P_{-}$) rotating clockwise
(anti-clockwise) around the $z$ axis. $P_{z}$ vanishes, for the
inversion asymmetry being exclusively along the $z$ direction.  At
$\vec{k}_{\parallel} = 0$, the states are degenerate and the ESP
becomes zero [$E_{+}(0) = E_{-}(0) = 0$ and $\vec{P}_{+} + \vec{P}_{-}
= 0$].

\subsection{Properties of the $L$-gap surface states}
\label{sec:dispersion.esp}
\subsubsection{Dispersion}
\label{sec:dispersion}
\begin{figure}
  \centering
  \includegraphics[scale = 1.5]{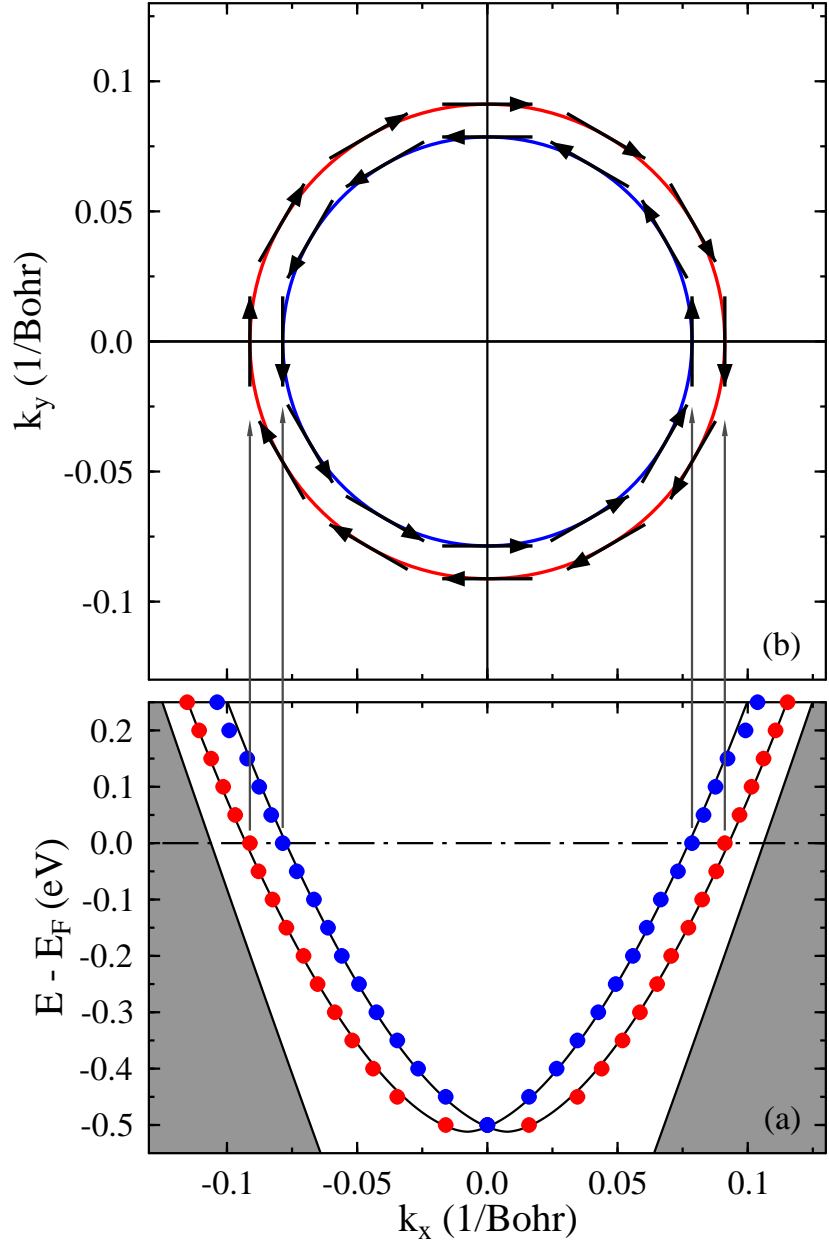}
  \caption{$L$-gap surface states on Au(111).
    (a) Dispersion of the spin-orbit split surface states along
    $\overline{\mathrm{K}}$--$\overline{\Gamma}$--$\overline{\mathrm{K}}$
    [i.\,e., $\vec{k}_{\parallel} = (k_{x}, 0)$]. Blue (red) symbols
    belong to the inner (outer) surface state. Grey arrows point from
    the surface states at the Fermi energy $E_{\mathrm{F}}$ to the
    momentum distribution shown in panel~b. The region of bulk bands
    is depicted by grey areas.  (b) Momentum distribution at
    $E_{\mathrm{F}}$. The thick arrows indicate the in-plane spin
    polarization [$P_{x}$ and $P_{y}$, according to
    eq.~(\ref{eq:1})].}
  \label{fig:dispersion}
\end{figure}

The dispersion of the surface states was obtained from the maxima in
the layer- and wavevector-resolved spectral density
[eq.~(\ref{eq:sd})] and is shown in Fig.~\ref{fig:dispersion}a.
The minimum energy is $-0.51~\mathrm{eV}$ which agrees well with the
experimentally observed value of $-0.49~\mathrm{eV}$ and with that of
an FLAPW (full-potential linearized augmented plane wave)
calculation.~\cite{Nicolay02} However, our theory gives a stronger
dispersion than the experiment: the Fermi wavenumbers
$k_{\mathrm{F}}^{\mathrm{in}} = 0.079 / \mathrm{Bohr}$ and
$k_{\mathrm{F}}^{\mathrm{out}} = 0.091 / \mathrm{Bohr}$ for the inner
and the outer surface states, resp., are slightly too small compared
to the experimental values of $0.091 / \mathrm{Bohr}$ and $0.104 /
\mathrm{Bohr}$ [FLAPW calculations with the WIEN code gave the same
trend.~\cite{Nicolay02} It can probably be related to the LDA surface
barrier which is in general to steep compared to an image-potential
barrier. The trend is furthermore consistent with energy-dependent
surface barriers that were introduced to describe correctly
image-potential states on Pd(110): these barriers became smoother with
increasing electron energy~\cite{Tamura90a}]. The spin-orbit splitting
(in $\vec{k}_{\parallel}$) agrees well: $0.012 / \mathrm{Bohr}$
(theory) to $0.013 / \mathrm{Bohr}$ (experiment), hence corroborating
that the important surface-states properties are well described by our
theory.

In semiconductor 2DEG's, the spin-splitting energy at zero magnetic
field is typically in the order of a few meV (e.\,g.,
$2.5$--$3.5~\mathrm{meV}$ from Refs.~\onlinecite{Luo88,Das90}). The
corresponding values for the spin-orbit coupling
$\gamma_{\mathrm{so}}$, as obtained from Shubnikov-de Haas
oscillations, range from about $0.7 \cdot 10^{-9}~\mathrm{eV\,cm}$
(Refs.~\onlinecite{Engels97,Schaepers98a}) to $0.9 \cdot
10^{-9}~\mathrm{eV\,cm}$ (Refs.~\onlinecite{Luo88,Luo90}).  The Rashba
effect appears to be considerably larger for the Au(111) surface: the
spin-splitting energy of about $0.14~\mathrm{eV}$ corresponds to a
$\gamma_{\mathrm{so}}$ of about $4.4 \cdot 10^{-9} \,\mathrm{eV\,cm}$.
More predicative in this context is the relative
$k_{\parallel}$-splitting $\Delta k_{\mathrm{F}} = 2
(k_{\mathrm{F}}^{\mathrm{out}} - k_{\mathrm{F}}^{\mathrm{in}}) /
(k_{\mathrm{F}}^{\mathrm{out}} + k_{\mathrm{F}}^{\mathrm{in}})$, which
equals to about $14~\%$ for Au(111).  For a 2DEG, this quantity is
directly related to the carrier densities of the spin-split states,
resulting in $\Delta k_{\parallel} \approx 4~\%$
(Ref.~\onlinecite{Luo88}). We note in passing that the splitting of
the $sp$-derived states on Au(111) is less than that for comparable
$d$-derived surface states on W(110),~\cite{Rotenberg99} possibly due
to the increase of SOC with angular momentum ($H_{\mathrm{SOC}}
\propto \vec{l} \cdot \vec{\boldsymbol{\sigma}}$).

In accord with eq.~(\ref{eq:3}), the momentum distributions at the
Fermi energy $E_{\mathrm{F}}$ are concentric circles
(Fig.~\ref{fig:dispersion}b), confirming the nomenclature of an
`inner' and an `outer' surface state (cf.\ also the constant-energy
cuts in Fig.~\ref{fig:rashba}).

\subsubsection{Spin polarization at the Fermi energy}
\label{sec:esp.ef}
Considering the point group $3m$ of the (111) surface and
time-reversal symmetry suggests that the leading terms of the electron
spin polarization (ESP) at a fixed energy can be written as
\begin{equation}
  \vec{P}(\varphi_{\mathrm{e}})
  =
  \left(
    \begin{array}{c}
       \alpha \sin\varphi_{\mathrm{e}}
      \\
      -\alpha \cos\varphi_{\mathrm{e}}
      \\
      \beta \cos 3\varphi_{\mathrm{e}}
    \end{array}
  \right)
  \label{eq:1}
\end{equation}
Hence, the spin polarization rotates clockwise (anti-clockwise) for
$\alpha > 0$ ($\alpha < 0$) around the $z$ axis (surface normal).
Evidently, the net spin polarization at the surface is zero and the
system remains nonmagnetic. Further, the signs for $\alpha$ of the two
spin-split surface states should be opposite,
$\mathrm{sgn}(\alpha^{\mathrm{in}}) =
-\mathrm{sgn}(\alpha^{\mathrm{out}})$ [Vanishing SOC requires that
$\vec{P}(\vec{k}_{\parallel}) = 0$ if summed over both states; that is
$\alpha^{\mathrm{in}} = -\alpha^{\mathrm{out}}$].  The nonzero $P_{z}$
reflects directly the threefold symmetry of the surface. In
particular, $|P_{z}|$ is largest at integer multiples of
$\varphi_{\mathrm{e}} = n \cdot 60^{\circ}$, $n$ integer, that is in
the directions of the first-nearest neighbor atoms within the surface
layer (Fig.~\ref{fig:geometry}). To our knowledge, the ESP, in
particular the modulus and signs of $\alpha^{\mathrm{in}}$ and
$\alpha^{\mathrm{out}}$ as well as of $\beta^{\mathrm{in}}$ and
$\beta^{\mathrm{out}}$, were not addressed in detail up to now, in
particular with respect to the 2DEG results
(\ref{sec:relation-l-gap}).  For the latter, we obtained
$\alpha^{\mathrm{in}} = +1$ ($+100~\%$) and $\alpha^{\mathrm{out}} =
-1$ ($-100~\%$) [eq.~(\ref{eq:pinner})].  Further,
$\beta^{\mathrm{in}}$ and $\beta^{\mathrm{out}}$ vanish.

The spin polarization of the surface states is due to the gradient of
the surface potential which plays the role of the inversion asymmetry
in a 2DEG\@. The $z$-derivative is much larger than the in-plane
derivatives that are related to the surface-potential corrugation.
Therefore, $|\alpha| \gg |\beta|$ is expected.  Indeed, the
spin-resolved spectral densities of the outermost Au layer at the
Fermi energy gave $\alpha^{\mathrm{in}} \approx -96.7~\%$ and
$\alpha^{\mathrm{out}} \approx +92.6~\%$, whereas $\beta^{\mathrm{in}}
\approx -1.4~\%$ and $\beta^{\mathrm{out}} \approx +1.3~\%$.
Comparing these results with eq.~(\ref{eq:pinner}) suggests that the
Rashba parameter $\gamma_{\mathrm{so}}$ is negative for the Au(111)
surface, since a positive $\gamma_{\mathrm{so}}$ corresponds to
$\alpha^{\mathrm{in}} > 0$ and $\alpha^{\mathrm{out}} < 0$. The large
in-plane spin polarization is consistent with spin-resolved
photoemission experiments on W(110)-($1\times 1$)H that report on
100~\%\ ESP, with regard to experimental resolution and statistics
(see Fig.~2 in Ref.~\onlinecite{Hochstrasser02}).

That the surface states are not fully spin-polarized, as is the case
for the 2DEG [cf.\ eq.~(\ref{eq:pinner})] is a further manifestation
of the crystal structure of the (111) surface.  In order to
investigate this finding, we concentrated on the
$\overline{\Gamma}$--$\overline{\mathrm{M}}$ direction (that is,
$\varphi_{\mathrm{e}} = 90^{\circ}$ to obtain $P_{z} = 0$) and scaled
the spin-orbit interaction (\ref{sec:ab-init-calc}). For vanishing
SOC, the surface-state wavefunctions are pure Pauli spinors and their
spatial parts are degenerate. Hence, $\alpha_{\mathrm{in}} = -
\alpha_{\mathrm{out}}$ and $\beta_{\mathrm{in}} = -
\beta_{\mathrm{out}}$, and the net ESP at a certain
$\vec{k}_{\parallel}$ consequently vanishes. With increasing SOC, and
hence increasing splitting, each wavefunction gets an admixture of the
other spin orientation. Furthermore, the spatial parts of the
wavefunctions are no longer degenerate. In other words, the difference
in $|\alpha^{\mathrm{in}}|$ and $|\alpha^{\mathrm{out}}|$ can be
attributed to the different `locations' in the two-dimensional
Brillouin zone (2BZ) of the SOC-split surfaces states.

\begin{figure}
  \centering
  \includegraphics[scale = 1.5]{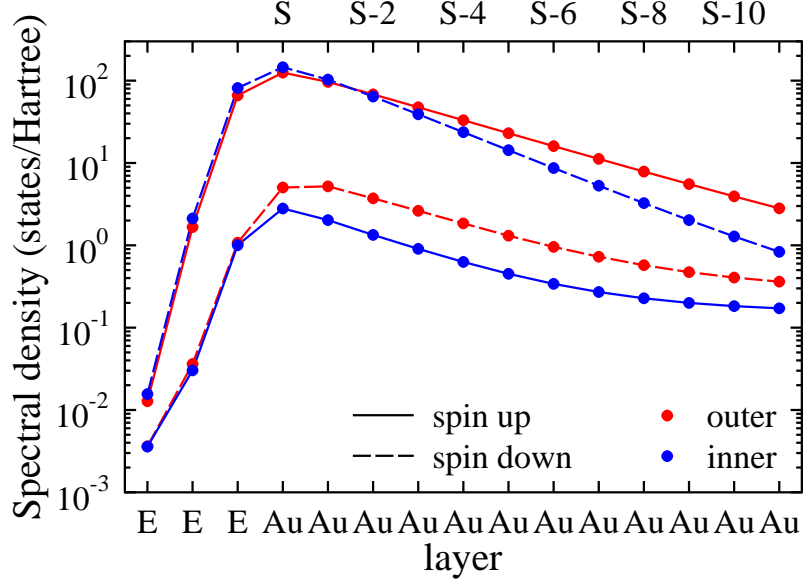}
  \caption{Layer- and spin-resolved spectral density of the surface states
    at the Fermi energy. The wavevectors are along
    $\overline{\Gamma}$--$\overline{\mathrm{M}}$ [$\vec{k}_{\parallel}
    = (0, k_{\mathrm{F}}^{\mathrm{in}})$ and $(0,
    k_{\mathrm{F}}^{\mathrm{out}})$, respectively]. ``E'' and ``Au''
    denote vacuum and Au layers, respectively. ``S'', ``S-2'' give the
    standard nomenclature for surface layers, starting with the
    outermost Au layer ``S''.  The terms ``spin up'' and ``spin down''
    refer to $P_{x}$ [cf.\ eq.~(\ref{eq:1}) with $\varphi_{\mathrm{e}}
    = 90^{\circ}$]. Note the logarithmic scale of the abscissa.}
  \label{fig:weight}
\end{figure}

This finding is supported by the layer- and spin-resolved spectral
density of the surface states integrated over the muffin-tin spheres
[eq.~(\ref{eq:sd})].  The spectral weight extends considerably into
the bulk (about 12 layers; Fig.~\ref{fig:weight}),
in agreement with recent calculations using a slab
geometry.~\cite{Nicolay02} The most striking fact, however, is that
both inner and outer surface states decay differently towards the bulk
and do not show full spin polarization (Note that the only nonzero ESP
component along $\overline{\Gamma}$--$\overline{\mathrm{M}}$ is
$P_{x}$, due to symmetry reasons). The spin polarization decreases (in
absolute value) towards the bulk, providing evidence for the surface
origin of the spin-orbit induced splitting.

Summarizing at this point, the properties of the $L$-gap surface
states show a close correspondence to those of the electronic states
in a 2DEG with Rashba spin-orbit interaction. The crystal structure of
the (111) surface leads in particular to a slightly reduced degree of
spin polarization and to a nonzero but small $P_{z}$. Further, crystal
properties show up in different degrees of localization for the inner
and the outer surface state.

\subsection{Probing the spin polarization by photoelectron spectroscopy}
\label{sec:probing}
\subsubsection{Spin-polarization effects in photoemission}
\label{sec:spe}
Depending on the photoemission set-up, one can easily produce
spin-polarized photoelectrons from a nonmagnetic surface, an effect
mainly due to spin-orbit coupling in the initial states. For
circularly polarized light, the effect is commonly termed `optical
orientation' because the photoelectron spin is aligned along the
incidence direction of the light while its orientation is determined
by the light helicity.~\cite{Woehlecke84} For linearly polarized
light, different effects were theoretically predicted and
experimentally confirmed (see references given
in~\ref{sec:phot-calc}).

The major aspect for producing spin-polarized photoelectrons from a
nonmagnetic surface is the symmetry of the entire set-up which
comprises the crystal surface, light polarization and incidence
direction as well as the electron-detection direction. As a rule of
thumb, one can assume that the less symmetry, the more components of
the photoelectron ESP are nonzero. In order to reliably probe the spin
polarization of an initial state (here: an $L$-gap surface state) one
has to assure that only those components of the photoelectron ESP are
nonzero that are also nonzero for the initial state. This restriction
implies that one has to choose the `correct' photoemission set-ups.
Otherwise, it might be difficult---if not impossible---to conclude
from the photoelectron ESP on that of the initial state. The main
complication in probing the ESP of the $L$-gap surface states arises
from the indispensable off-normal emission of the photoelectrons. It
reduces the symmetry considerably (compared to normal emission) and,
hence, allows for more $\vec{P}$-components being
nonzero.~\cite{Henk98a}

In the following, two main types of photoemission set-ups will be
discussed. In the first one, $\vec{k}_{\parallel}$ lies in a mirror
plane of the surface, i.\,e., $\vec{k}_{\parallel}$ along
$\overline{\mathrm{M}}$--$\overline{\Gamma}$--$\overline{\mathrm{M}}$
(cf.\ Fig.~\ref{fig:geometry}b). Since for $\varphi_{\mathrm{e}} =
90^{\circ}$ and $180^{\circ}$ [$\vec{k}_{\parallel} = (0, k _{y}$)],
the initial-state ESP is aligned along $x$ [eq.~(\ref{eq:1})], the
light has to be chosen in such a way that the mirror operation $x \to
-x$ is retained.  For the second type, $\vec{k}_{\parallel}$ is along
$\overline{\mathrm{K}}$--$\overline{\Gamma}$--$\overline{\mathrm{K}}$
[$\vec{k}_{\parallel} = (k _{x}, 0$)] and only the trivial symmetry
operation remains ($\vec{k}_{\parallel}$ perpendicular to a mirror
plane). Therefore, it is not possible to choose incidence direction
and polarization of the light in such a way that only $P_{y}$ is
nonzero.

The following results were obtained for linearly and circularly
polarized light with photon energy $\omega = 21.22~\mathrm{eV}$
(He$_{\mathrm{I}}$) incident at a polar angle $\vartheta_{\mathrm{ph}}
= 45^{\circ}$. Fixing the initial-state energy at $E_{\mathrm{F}}$, we
are concerned with constant initial-energy spectroscopy (CIS).  Our
results hold qualitatively also for other parameters (e.\,g., polar
angle of incidence, initial-state energy, and photon energy).

\subsubsection{$\overline{\Gamma}$--$\overline{\mathrm{M}}$}
\label{sec:gm}
For
$\overline{\mathrm{M}}$--$\overline{\Gamma}$--$\overline{\mathrm{M}}$,
one obtains from eq.~(\ref{eq:1}) that the spin polarization of the
initial state is aligned along the $x$ axis, $\vec{P} = (\alpha, 0,
0)$. For off-normally incident p-polarized light, the photoelectron
spin polarization is normal to the scattering plane [see
Ref.~\onlinecite{Tamura91b} for (001) surfaces].  Hence, to probe the
initial-state spin polarization for
$\overline{\mathrm{M}}$--$\overline{\Gamma}$--$\overline{\mathrm{M}}$
($\varphi_{\mathrm{e}} = 90^{\circ}$) one chooses a light incidence
within the $yz$ plane ($\varphi_{\mathrm{ph}} = 90^{\circ}$ or
$270^{\circ}$) which produces a nonzero $P_{x}$ only.

\begin{figure}
  \centering
  \includegraphics[scale = 1.5]{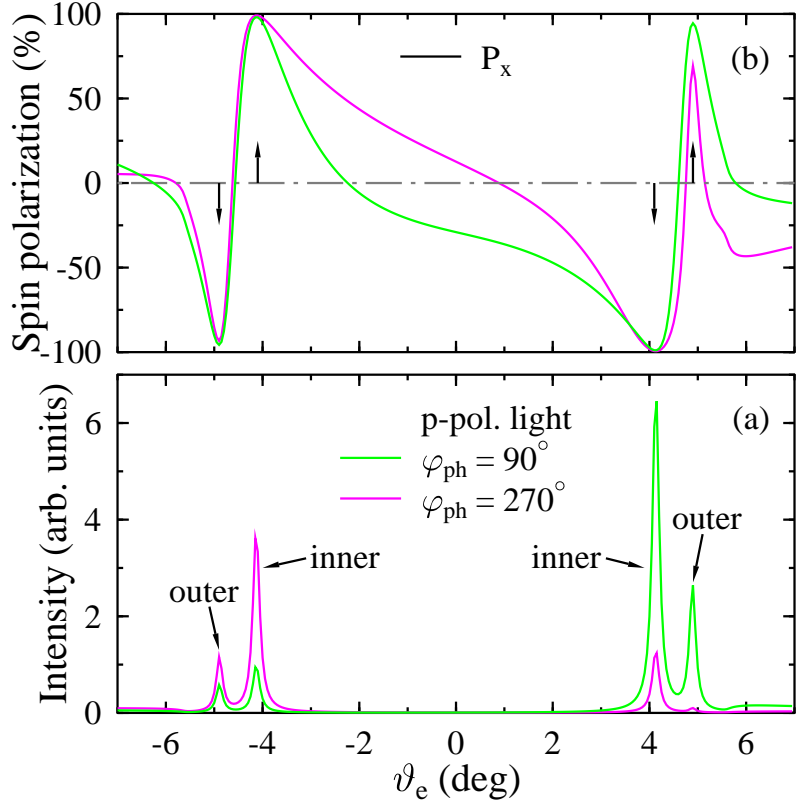}
  \caption{Spin-resolved photoemission for $\vec{k}_{\parallel}$ along
    $\overline{\mathrm{M}}$--$\overline{\Gamma}$--$\overline{\mathrm{M}}$
    ($\varphi_{\mathrm{e}} = 90^{\circ}$) and p-polarized light
    ($\omega = 21.22~\mathrm{eV}$, $\vartheta_{\mathrm{ph}} =
    45^{\circ}$). (a) Intensities for $\varphi_{\mathrm{ph}} =
    90^{\circ}$ (green) and $270^{\circ}$ (magenta) azimuth of light
    incidence.  The surface-state maxima are indicated (`inner',
    `outer'). (b) Associated photoelectron spin polarization along
    $\vec{x}$. The vertical arrows mark the positions of the surface
    states.}
  \label{fig:p.gm}
\end{figure}

Scanning the polar angle of emission $\vartheta_{\mathrm{e}}$
(Fig.~\ref{fig:p.gm}a), the intensities show four distinct maxima which
lie symmetrically around $\overline{\Gamma}$ ($\vec{k_{\parallel}} =
0$ or $\vartheta_{\mathrm{e}} = 0^{\circ}$).
The inner surface state appears at $\vartheta_{\mathrm{e}} = \pm
4.2^{\circ}$, the outer at $\vartheta_{\mathrm{e}} = \pm 4.9^{\circ}$.
Bulk contributions to the photocurrent occur for $|
\vartheta_{\mathrm{e}}| > 6^{\circ}$, as can be seen by the very small
intensities (compared to the surface-state intensities).

The $x$-component of the photoelectron ESP shows distinct minima and
maxima at the positions of the surface states (Fig.~\ref{fig:p.gm}b).
The sign of $P_{x}$, and hence the sign of $\alpha$
[eq.~(\ref{eq:1})], corresponds to those obtained from the
spectral-density calculations for the initial state.  Even the
magnitudes agree well: from Fig.~\ref{fig:p.gm} one would deduce
$\alpha^{\mathrm{in}} \approx -99~\%$ and $\alpha^{\mathrm{out}}
\approx +93~\%$, compared to $\alpha^{\mathrm{in}} \approx -97~\%$ and
$\alpha^{\mathrm{out}} \approx +93~\%$ for the initial states. That
intensities and spin polarizations for $\vartheta_{\mathrm{ph}} =
90^{\circ}$ and $270^{\circ}$ as well as for
$\pm\vartheta_{\mathrm{e}}$ differ is attributed to the
transition-matrix elements which obviously depend on the direction of
light incidence [Note in this context the ABC stacking sequence of the
(111) surface, Fig.~\ref{fig:geometry}].

\begin{figure}
  \centering
  \includegraphics[scale = 1.5]{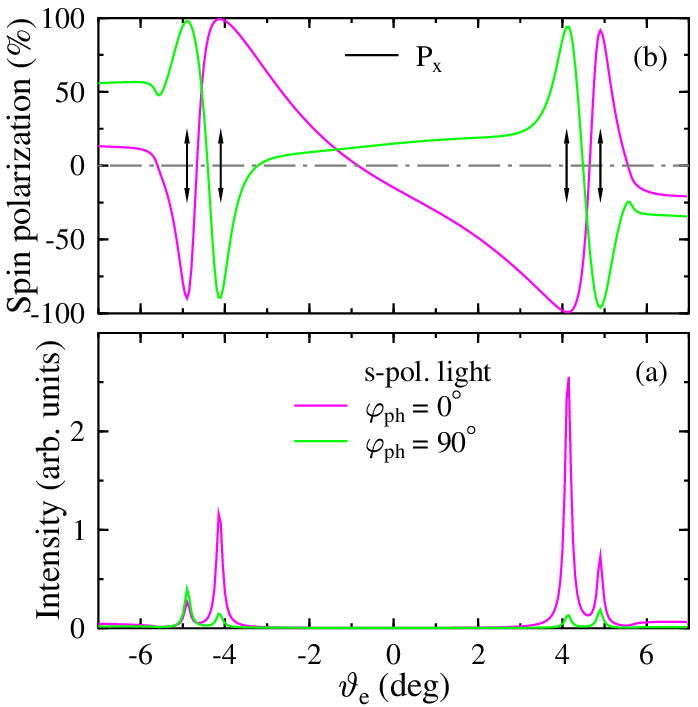}
  \caption{As Fig.~\ref{fig:p.gm} but for s-polarized light
    ($\omega = 21.22~\mathrm{eV}$).  (a) Intensities for
    $\varphi_{\mathrm{ph}} = 0^{\circ}$ (magenta) and $90^{\circ}$
    (green) azimuth of light incidence. (b) Associated photoelectron
    spin polarization along $\vec{x}$.}
  \label{fig:s.gm}
\end{figure}

The use of p-polarized light nicely provides access to the spin
polarization of the initial state. For s-polarized light, however,
this is not completely true, as can be seen in Fig.~\ref{fig:s.gm}.
In this case, $P_{x}$ is the only nonzero ESP component, too.  For
$\varphi_{\mathrm{ph}} = 0^{\circ}$ (that is, for the electric-field
vector $\vec{E}$ of the incident radiation parallel to $\vec{y}$),
$P_{x}$ shows the same structure as for p-polarized light (magenta in
Fig.~\ref{fig:s.gm}b). However, for $\varphi_{\mathrm{ph}} =
90^{\circ}$ ($\vec{E} \parallel \vec{x}$), one observes the opposite
behavior: a positive $\alpha^{\mathrm{in}}$ and a negative
$\alpha^{\mathrm{out}}$ (green in Fig.~\ref{fig:s.gm}b). This finding
is a direct manifestation of spin-orbit coupling (SOC).  Without SOC,
an initial-state wavefunction would be either even or odd under the
mirror operation $x \to -x$. The spatial parts for `spin up'
($\uparrow$) and `spin down' ($\downarrow$) would be identical, giving
rise to an unpolarized state. However, spin-orbit coupling mixes even
and odd initial-state wavefunctions.~\cite{Henk96a} Schematically, one
can write for the initial-state wavefunction
\begin{equation}
  \mid \Psi \rangle
  =
  \mid \Psi_{\mathrm{even}} \rangle \chi^{\tau}
  +
  \mid \Psi_{\mathrm{odd}} \rangle \chi^{-\tau},
  \quad
  \tau = \uparrow, \downarrow.
\end{equation}
In the dipole approximation, $\vec{E}$ couples to the even spatial
part of the initial-state wavefunction if lying in the mirror plane.
This is the case for p-polarized light as in Fig.~\ref{fig:p.gm} or
for s-polarized light with $\varphi_{\mathrm{ph}} = 0^{\circ}$. It
couples to the odd part if being perpendicular, as is the case for
s-polarized light incident at $\varphi_{\mathrm{ph}} = 90^{\circ}$.
Hence, one can conclude that the even parts of the initial states are
dominant (cf.\ the intensities in Fig.~\ref{fig:s.gm}: large for
$\varphi_{\mathrm{ph}} = 0^{\circ}$, small for $\varphi_{\mathrm{ph}}
= 90^{\circ}$; this is confirmed by the angular momentum- and
spin-resolved spectral densities) and produce a negative
$\alpha^{\mathrm{in}}$ and a positive $\alpha^{\mathrm{out}}$. The
intensity difference for $\pm \vartheta_{\mathrm{e}}$ can again be
attributed to the fcc lattice which is not symmetric with respect to
the $xz$ plane (ABC stacking sequence).

\begin{figure}
  \centering
  \includegraphics[scale = 1.5]{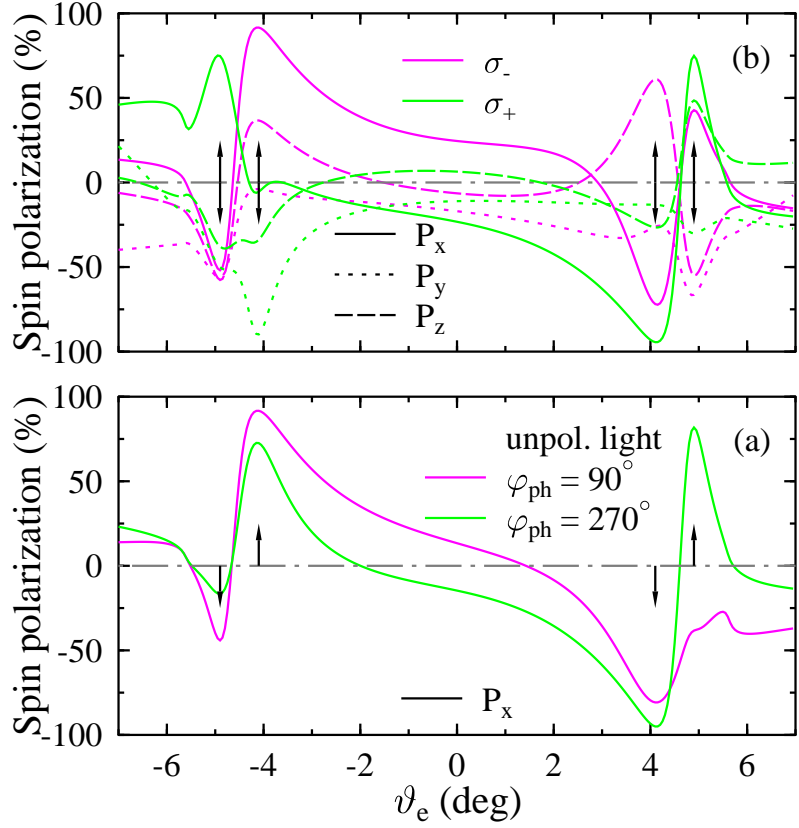}
  \caption{Spin-resolved photoemission for $\vec{k}_{\parallel}$ along
    $\overline{\mathrm{M}}$--$\overline{\Gamma}$--$\overline{\mathrm{M}}$
    ($\varphi_{\mathrm{e}} = 90^{\circ}$, $\omega =
    21.22~\mathrm{eV}$, $\vartheta_{\mathrm{ph}} = 45^{\circ}$).  The
    vertical arrows mark the surface-state positions. (a)
    Photoelectron spin polarization along $x$ for p-polarized light
    incident at $\varphi_{\mathrm{ph}} = 90^{\circ}$ (magenta) and
    $270^{\circ}$ (green) azimuth.  (b): As panel~a, but for
    circularly polarized light for positive ($\sigma_{+}$, green) and
    negative ($\sigma_{-}$, magenta) helicity incident at
    $\varphi_{\mathrm{ph}} = 0^{\circ}$.  The components of the
    photoelectron spin polarization are differentiated by line
    styles.}
  \label{fig:uc.gm}
\end{figure}

Unpolarized light can be regarded as an incoherent superposition of s-
and p-polarized light. Since one detects with s-polarized light
incident within the $yz$ plane ($\varphi_{\mathrm{ph}} = 90^{\circ}$)
the `wrong' ESP (green in Fig.~\ref{fig:s.gm}b), a question arises
whether unpolarized light provides nevertheless the `correct'
initial-state ESP.  Indeed, $P_{x}$ shows the $-/+$--$+/-$ shape
(Fig.~\ref{fig:uc.gm}a), but the ESP is significantly reduced in
absolute value.
The latter can be explained by the smaller intensity for s-polarized
light than for p-polarized light. Hence, using a He$_{\mathrm{I}}$
rare-gas discharge lamp provides information on the sign but not on
the magnitude of the initial-state spin polarization.

In order to probe the initial-state $\vec{P}$ with `optical
orientation', one would choose circularly polarized light incident at
$\varphi_{\mathrm{ph}} = 0^{\circ}$ or $180^{\circ}$, expecting that
mainly $P_{x}$ would be produced, although all three components of
$\vec{P}$ become nonzero for $\vartheta_{\mathrm{ph}} \not=
0^{\circ}$.  In this case, the intensities $I$ and spin polarizations
$\vec{P}$ obey $(I, P_{x}, P_{y}, P_{z}) \to (I, P_{x}, -P_{y},
-P_{z})$ when changing $\varphi_{\mathrm{ph}}$ from $0^{\circ}$ to
$180^{\circ}$ and simultaneously reversing the light helicity
$\sigma_{\pm}$ [$I(\sigma_{+}) \not= I(\sigma_{-})$ means that there
is circular dichroism in angular distribution (CDAD)].  As is evident
from Fig.~\ref{fig:uc.gm}b, the photoelectron ESP shows a complicated
behavior, from which it is almost impossible to conclude on the
initial-state spin polarization without prior (theoretical) knowledge.

\subsubsection{$\overline{\Gamma}$--$\overline{\mathrm{K}}$}
\label{sec:gk}
With $\vec{k}_{\parallel}$ along
$\overline{\mathrm{K}}$--$\overline{\Gamma}$--$\overline{\mathrm{K}}$
($\varphi_{\mathrm{e}} = 0^{\circ}$) the only remaining symmetry
operation is the trivial one, which yields that all components of the
photoelectron ESP are generally nonzero. The initial-state ESP,
however, reads $\vec{P} = (0, -\alpha, \beta)$. Thus, a nonzero
$P_{x}$ of the photoelectron would be a direct manifestation of a ESP
due to the photoemission process.  We have performed photoemission
calculations for different light polarizations and incidence
directions and found that in most cases the photoelectron spin
polarization is hardly to relate to that of the initial state.  The
most promising results were obtained for p-polarized light incident in
the $xz$ plane ($\varphi_{\mathrm{ph}} = 0^{\circ}$ and
$180^{\circ}$). The symmetry of this set-up implies certain relations
between the intensities $I$ and the photoelectron spin polarization
$\vec{P}$: changing simultaneously $\varphi_{\mathrm{ph}}$ from
$0^{\circ}$ to $180^{\circ}$ and $\vartheta_{\mathrm{e}}$ to
$-\vartheta_{\mathrm{e}}$ results in $(I, P_{x}, P_{y}, P_{z}) \to (I,
P_{x}, -P_{y}, -P_{z})$, that is, $P_{y}$ and $P_{z}$ change sign
whereas $I$ and $P_{x}$ remain unaffected. Therefore, it is sufficient
to discuss only the case $\varphi_{\mathrm{ph}} = 0^{\circ}$.

\begin{figure}
  \centering
  \includegraphics[scale = 1.5]{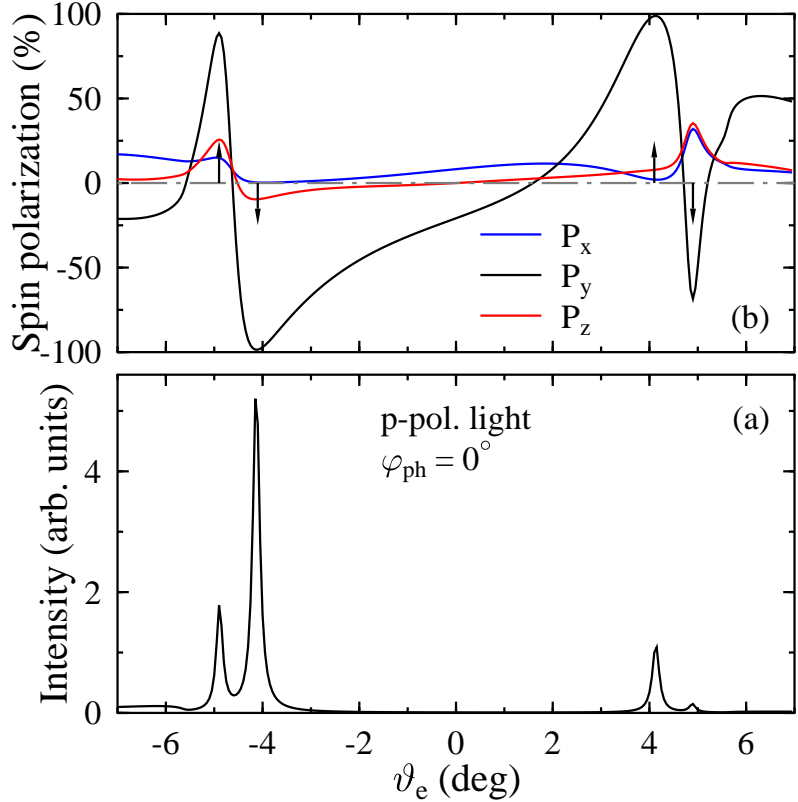}
  \caption{Spin-resolved photoemission for $\vec{k}_{\parallel}$ along
    $\overline{\mathrm{K}}$--$\overline{\Gamma}$--$\overline{\mathrm{K}}$
    ($\varphi_{\mathrm{e}} = 0^{\circ}$) and p-polarized light
    ($\omega = 21.22~\mathrm{eV}$, $\vartheta_{\mathrm{ph}} =
    45^{\circ}$). (a) Intensities for $\varphi_{\mathrm{ph}} =
    0^{\circ}$ azimuth of light incidence. (b) Associated
    photoelectron spin polarization.  The vertical arrows mark the
    surface-state positions.}
  \label{fig:p.gk}
\end{figure}

As is evident from Fig.~\ref{fig:p.gk}, the intensity maxima occur at
the same polar angles of emission as along
$\overline{\mathrm{M}}$--$\overline{\Gamma}$--$\overline{\mathrm{M}}$
which proves the circular shape of the momentum distribution
(Fig.~\ref{fig:dispersion}b).
The dominant component of $\vec{P}$, $P_{y}$ (black in
Fig.~\ref{fig:p.gk}b), shows a $+/-$-$+/-$ shape which agrees well
with that of the initial-state spin polarization
(Fig.~\ref{fig:dispersion}b).  Further, one finds
$|\alpha^{\mathrm{in}}| > |\alpha^{\mathrm{out}}|$. However, to
conclude on the magnitude on $\alpha$ is rather difficult, in
particular for $\alpha^{\mathrm{out}}$ because of the sizable $P_{x}$
and $P_{z}$. These $\vec{P}$ components are rather large, especially
for the outer surface state on which we will focus now. The comparison
of the $P_{z}$ values of $25.7~\%$ and $35.3~\%$ at
$\vartheta_{\mathrm{e}} = -4.9^{\circ}$ and $+4.9^{\circ}$ (red in
Fig.~\ref{fig:p.gk}b), respectively, with those of the initial state
($-1.3~\%$ and $+1.3~\%$) renders it impossible to conclude from the
photoelectron spin polarization on that of the surface state. The same
holds for $P_{x}$ ($14.8~\%$ and $32.0~\%$; blue in
Fig.~\ref{fig:p.gk}b) which is exclusively due to the photoemission
process.

Despite this negative but expected result, one can speculate to probe
the initial-state $P_{z}$ by altering the system.  First, one needs a
spin-polarization effect which produces a dominant $P_{z}$. This could
be accomplished by `optical orientation' with normally incident
circularly polarized light, but for off-normal emission the other
$\vec{P}$ components are too large to conclude undoubtedly on $\beta$.
Second, since the small initial-state $P_{z}$ arises from the
corrugation of the surface potential, one might think about increasing
the corrugation by covering the surface by an adlayer, e.\,g.,
$(\sqrt{3} \times \sqrt{3})\mathrm{R}30^{\circ}$-Xe/Au(111). That the
splitting of the surface states can indeed be changed by coverage was
already found for Li/W(110).~\cite{Rotenberg99}

Summarizing, the photoemission calculations prove that it is possible
to conclude from the photoelectron spin polarizations on those of the
initial $L$-gap surface states, provided the set-up is chosen
correctly. Otherwise, the photoelectron spin polarization which is
brought about by the photoemission process itself obscures the
property of interest.

\section{Concluding remarks}
\label{sec:conclusion}
Our theoretical investigations reveal on one hand a striking
similarity between the electronic states in a two-dimensional electron
gas (2DEG) with Rashba spin-orbit interaction and the $L$-gap surface
states on Au(111). On the other hand, the structure of the Au(111)
surface produces a nonzero but small spin-polarization component
normal to the surface that is missing in a 2DEG\@. To probe the spin
polarization of the spin-orbit split surface states by spin- and
angle-resolved photoelectron spectroscopy can completely fail if the
set-up is badly chosen. As a rule of thumb, those set-ups work best
that produce a photoelectron spin polarization aligned along that of
the initial state (see, e.\,g., Refs.~\onlinecite{Feder96} and
\onlinecite{Henk96a}). We would like to encourage strongly experiments
in order to confirm our theoretical results.

The $L$-gap surface states can be regarded as a source for highly
spin-polarized electrons with unique properties. Hence, one can
speculate whether the Au(111) surface can be used as a model system
for spintronics if brought into contact with magnetic material.

%
%
\bibliography{short,refs,refs2,refs3}

\begin{thebibliography}{47}
\bibitem[{Uni()}]{Units}
\bibinfo{note}{We use atomic Hartree units, $e = \hbar = m = 1$, $c \approx
  137.036$.}
\end{thebibliography}
\bibliographystyle{apsrev}

\end{document}